\documentclass[aps,twocolumn,a4paper]{revtex4}
\usepackage{graphicx}
\usepackage{amsmath,amssymb}
\usepackage{enumerate}
\usepackage{subfigure}
\usepackage{tabularx}
\usepackage{subfigure}  
\usepackage{xcolor}

\newcommand{\be}{\begin{equation}}
\newcommand{\ee}{\end{equation}}
\newcommand{\ben}{\begin{eqnarray}}
\newcommand{\een}{\end{eqnarray}}
\newcommand{\bes}{\begin{subequations}}
\newcommand{\ees}{\end{subequations}}

\newcommand{\bb}{\bibitem}

\begin{document}
\title{ Configurational entropy of skyrmions and half-skyrmions in planar magnetic elements}
\author{D. Bazeia}
\affiliation{Departamento de F\'\i sica, Universidade Federal da Para\'\i ba, 58051-970 Jo\~ao Pessoa, PB, Brazil}
\author{ E. I. B. Rodrigues}
\affiliation{Universidade Federal Rural de Pernambuco, 54518-430 Cabo de Santo Agostinho, PE, Brazil}

\begin{abstract}
This works deals with the presence of localized planar structures in magnetic materials that admit integer or half-integer topological charge. We study models in which the internal disposition of magnetization is driven by a single parameter, that controls the topological charge density of the magnetic structure. In particular, we focus mainly on the configurational entropy of these skyrmions and half-skyrmions and show how to increase or diminish their informational capabilities.
\end{abstract}
\maketitle
\section{Introduction}

Topological structures appear in several areas of physics. In high energy physics, there are kinks in the real line, vortices in the plane and monopoles in the tridimensional space; see, e.g., Refs. \cite{B1,B2,B3}. In condensed matter, there are magnetic materials that support skyrmions and half-skyrmions; see, e.g., Refs. \cite{S0,S1,S2,S3,S4,S4a,S5,nature1,nature2} and references therein. Other possibilities have appeared in \cite{024415,18804} and in \cite{224407,5603}, and they are also studied in multilayered elements \cite{S5,177201,yu2018}, in Bose-Einstein condensates \cite{Yang} and in liquid crystals \cite{062706}. For a recent review on skyrmions in condensed matter, see also Ref. \cite{BS}.

Skyrmions appeared before in \cite{Sa,Sb}, in the context of hadron physics, where a meson model could support topological solutions with integer topological charge to be associated to baryons \cite{Sc}. In the present work, however, we want to study magnetic skyrmions and half-skyrmions, so we turn attention to planar magnetic structures. An alternative to investigate localized structures of the skyrmion and half-skyrmion type was developed in \cite{BDR,JMM1,JMM2}, where the magnetic structures are constructed via analytical solutions of planar relativistic scalar field models \cite{PRL}. We follow this route to investigate three distinct models that support skyrmions with unity topological charge, and three others models that support half-skyrmions with topological charge one-half.  

Related to localized structures, the concept of configurational entropy was introduced for relativistic scalar fields in \cite{GS}, based on the Shannon's mathematical theory of communication \cite{SE}. This conformational entropy (CE) has been studied in several contexts, for instance, within the AdS/QCD correspondence \cite{762,776,787,786,106002}, in gravity \cite{044046,552,083509,381,772}, topological defects \cite{755,1970035}, and more recently in the context of skyrmions in \cite{JMM3,174440}. In \cite{JMM3}, in particular, the CE of skyrmions was investigated for the first time, and in \cite{174440}, another route to investigate the CE for skyrmions was developed, with the result that the CE can help us understand the thermal fluctuations of the skyrmion energy. See also \cite{Inf,New} for other related investigations.    

In the present work, we focus on the calculation of the CE for skyrmions and half-skyrmions in several distinct models. In order to implement the investigation, we organize the work as follows: in the next Sec. II we introduce the basic tools to deal with skyrmions and half-skyrmions, following the lines that appeared before in Refs. \cite{BDR,JMM1,JMM2}, and to investigate the CE, as presented in \cite{JMM3}. With the theoretical ingredients described in Sec. II, we investigate in Sec. III the CE associated to six distinct models, three describing skyrmions, and three describing half-skyrmions. In Sec. IV we end the work with some comments and conclusions.

\section{Generalities} 

\subsection{ Skyrmions and Half-skyrmions} 

The standard procedure to describe the topological behavior of planar skyrmions in magnetic materials follows with the introduction of the quantity
\be\label{tc}
Q=\frac{1}{4\pi}\int \!\!\! \int dxdy \ {\bf m}\cdot\left(\frac{\partial{\bf m}}{\partial x}\times\frac{\partial{\bf m}}{\partial y}\right),
\ee
where ${\bf m}$ obeys ${\bf m}\cdot {\bf m}=1$ and is defined by ${\bf m}={\bf M}/{|\bf M|}$, where ${\bf M}$ is the magnetization of the magnetic material. The above quantity $Q$ is known as the topological charge of the magnetic structure: it is zero if ${\bf m}$ is uniformly distributed along the positive or negative ${\hat z}$ direction that defines the $(x,y)$ plane that describes the magnetic system. However, it may also describe two interesting distinct families of nontrivial configurations, one with integer charge $Q=\pm1$, and the other with half-integer charge $Q=\pm 1/2$.

In this work we shall suppose that the localized planar structures engender rotational symmetry along the ${\hat z}$ axis, so we work with cylindrical coordinates $({\hat r},{\hat \theta}, {\hat z})$. We then suppose that the magnetization only depends on the radial coordinate, such that ${\bf m}={\bf m}(r)$. Moreover, we also consider the case of helicoidal excitations, with the magnetization being a bidimensional vector ortogonal to the radial direction, obeying ${\bf m}\cdot {\hat r}=0$. As it is usually considered in the related literature, when $Q$ is integer, we refer to the magnetic structure as skyrmion, and for $Q$ being half-integer, to half-skyrmion. The fact that ${\bf m}$ obeys ${\bf m}\cdot{\bf m}=1$, ${\bf m}\cdot{\hat r}=0$, and ${\bf m}={\bf m}(r)$ allows that we write the magnetization in the form
\be\label{M}
{\bf m}(r)=\hat{\theta}\cos\Theta(r) + \hat{z}\sin\Theta(r).
\ee 
As we considered in Refs. \cite{BDR,JMM1,JMM2}, here we also take $\Theta(r)$ as the single degree of freedom to describe the magnetic behavior inside the planar magnetic material. Moreover, we define 
\be\label{T}
\Theta(r)=\frac{\pi}2\phi(r)+\delta,
\ee
where $\phi$ is a real scalar field and $\delta$ a constant phase that can be used to set the value of the magnetization at the origin, that is, at the center of the magnetic structure.  

In terms of the planar coordinates $r$ and $\theta$, we can rewrite Eq.~\eqref{tc} in the form
\be
Q=\int_{0}^\infty\!\!\! dr\; q(r),
\ee
where $q(r)$ is the density of topological charge, given by
\be\label{T}
q(r)=-\frac{\pi}{4} \frac{d\phi(r)}{dr}\cos\Theta(r).
\ee
We use this to write the topological charge in the form
\be\label{Q}
Q=\frac12 \left(\sin\Theta(0) - \sin\Theta(\infty)\right).
\ee
This result shows that the topological profile of the magnetic structure is directly related to the value of $\Theta(r)$ at the origin $r=0$, and asymptotically, as $r$ increases to larger and larger values.  As considered before in \cite{BDR,JMM1,JMM2}, here we also suppose that the scalar field $\phi$ is homogeneous and dimensionless, described by the the Lagrangean density \cite{PRL}
\be\label{model}
{\cal L}=\frac12\partial_\mu\phi\partial^\mu\phi-U(x_\mu x^\mu,\phi),
\ee
where
\be
\partial_\mu\phi=\frac{\partial\phi}{\partial x^\mu},
\ee
and $x^\mu=(x^0=t,x^1=x,x^2=y)$ is the spacetime position vector in Cartesian coordinates. Since we are searching for time-independent field configuration that engenders rotational symmetry, it is then supposed that $\phi = \phi(r)$, and that $U=U(r^2,\phi)$ has the form
\be\label{modelU}
U(r^2,\phi)=\frac1{r^2}\;P(\phi),
\ee
where $P(\phi)$ is in principle a smooth function of the field $\phi$. In this case, the equation of motion becomes
\be\label{modeleq}
r^2\frac{d^2\phi}{dr^2} + r\frac{d\phi}{dr} -\frac{d P}{d\phi} = 0.
\ee
A step of interest here is to recognise that solutions of the first order equations 
\be\label{ordeq}
r\frac{d\phi}{dr} = \pm \sqrt{2P(\phi)},
\ee
also solve the equation of motion \eqref{modeleq}, which is second order ordinary differential equation. This is important since the first order equations help us to find solutions and to show they are stable against small radial fluctuations. 

We can follow two distinct routes to study stability of the solutions: first, we consider the static solution $\phi(r)=\phi_s(r)+ \epsilon\;\eta(r)$, where $\phi_s(r)$ is the radial solution, $\epsilon$ is a small parameter and $\eta(r)$ is the radial fluctuation. We use $\phi(r)$ in the total energy, and expand it in terms of $\epsilon$  to show that $E_0$, the contribution to the energy at zero order in $\epsilon$, is the minimum energy. The second possibility is to write the dynamical field in the form $\phi(r,t)=\phi_s(r)+ \eta(r,t)$, suppose that $\eta(r,t)$ is small, use this in the equation of motion and expand it until first order in $\eta(r,t)$ and investigate how $\eta(r,t)$ evolves in time. The first possibility was used in Refs. \cite{BDR,JMM1,JMM2}, and the second one was used in \cite{JMM3}, so we do not include stability in the work. However, we would like to reenforce that skyrmions and half-skyrmions and also protected topologically. To see this, we notice that we are working in the three-dimensional space-tine, with time and $(x,y)$ or $(r, \theta)$. In this case, we can introduce the topological current density
\be
J^\mu=\frac{1}{8\pi} \varepsilon^{\mu\nu\lambda}\;\;  {\bf m}\cdot \partial_{\nu}{\bf m} \times \partial_{\lambda}{\bf m},
\ee
where $\varepsilon^{\mu\nu\lambda}$ is the Levi-Civita symbol. This current density is conserved, that is, $\partial_\mu J^\mu=0$. Thus, the charge density 
\be  
J^0=q=\frac{1}{4\pi}\;{\bf m}\cdot\left(\frac{\partial{\bf m}}{\partial x} \times \frac{\partial{\bf m}}{\partial y}\right),
\ee 
can be integrated to give the topological charge \eqref{tc}, which is topologically protected.

\subsection{Configurational Entropy} 

Based on Shannon's work on information \cite{SE}, Gleiser and Stamatopoulos introduced in Ref. \cite{GS} the concept of configuration entropy (CE), used to provide a connection between an entropic measure in the functional space and the energy of spatially localized configurations. In the present work, however, we want to focus on the topological behavior of the spatially localized magnetic configurations, so instead of considering the energy density, we use the topological charge density presented in Eq. \eqref{T}. The approach here is similar to the one recently considered in \cite{JMM3}. In this context, to investigate the CE associated to the several magnetic skyrmion-like configurations to be introduced below, we first calculate the Fourier transform of the topological charge density. In the $(r,\theta)$ plane, the Fourier transform $\sigma(k)$ has the form 
\be
\sigma( k) =\int^\infty_0 r\,dr\, q(r)\;J_0(kr) ,
\ee
where $J_0(kr)$ is Bessel function of zeroth order. We follow \cite{GS} and define the normalised modal fraction that accounts for the correlation distribution in momentum space; it has the form
\be\label{fm}
f(k)\equiv \frac{|\sigma(k)|^2}{\int d^{2} { k} |\sigma (k)|^2}.
\ee
We use this to define the CE as follows 
\be\label{sk}
S_C[f]\equiv \int d^{2} k\, {\cal D}_f(k) ,
\ee
where
\be\label{dsk}
{\cal D}_f(k) =-\widetilde{f}(k)\;\text{ln}[\widetilde{f}(k)],
\ee
and  $\widetilde{f}(k) = f(k)/f_{\text{max}}( k)$ is now limited:  $\widetilde{f}(k)\in [0,1]$, since $f_{\text{max}}( k)$ is the maximum value of $f(k)$. As we see, according to the Shannon's information theory, ${\cal{D}}_f(k)$ represents the CE density.

Although we are dealing with continuous systems, with an uncoutable number of states, let us illustrate the situation with a system composed of a given number of states. As a simple possibility, we take an ideal dice with six equally probable faces. As one knows from Shannon's work \cite{SE}, if one paints each one of the six faces with red, blue, green, orange, yellow and violet, the associated Shannon entropy is $\ln(6)$. However, if one takes another ideal dice and paint two faces red, two blue and the other two green, the Shannon entropy is now $\ln(3)$. Moreover, if the dice has three faces painted red, and the other three painted blue, the Shannon entropy becomes $\ln(2)$. In the simple dice system, one notices that the Shannon entropy decreases as we decrease the number of distinct possible configurations, and this will help us understand the results associated to the CE of skyrmions and half-skyrmions that we study in this work.

\begin{figure}[t!]
\centering
\includegraphics[scale=0.57]{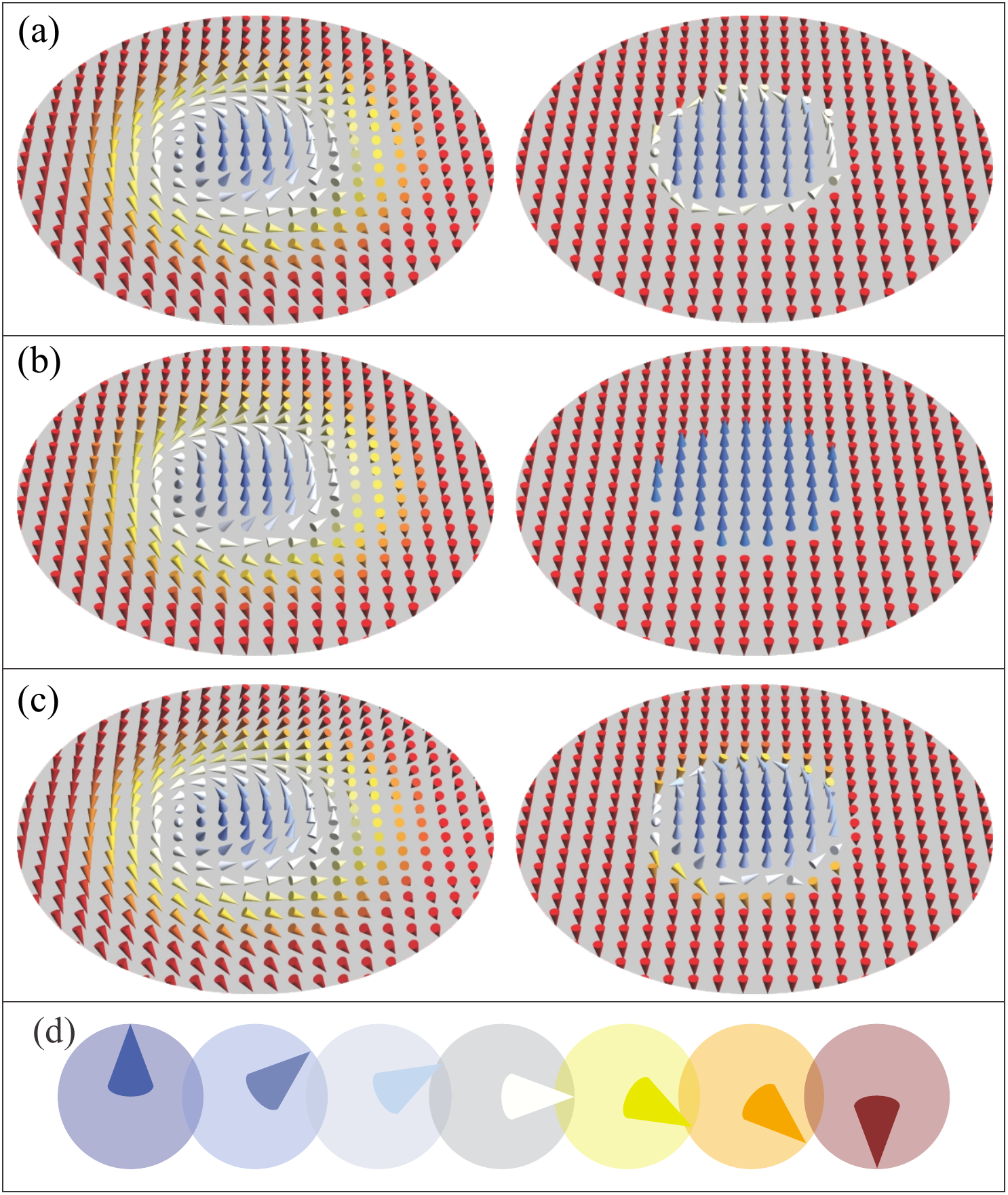}
\caption{The magnetic structures with $Q=1$ for (a) model \eqref{p1}, (b) model \eqref{p2} and (c) model \eqref{p5}, depicted for $s=0.3$ (left) and $s=0.9$ (right). The bottom panel in (d) illustrates how the magnetization ${\bf m}(0)=\hat{z}$ (blue) changes to $-\hat{z}$ (red) as $r$ increases to larger and larger values.}
\label{skyrmion}
\end{figure}

\section{Models} 

Let us now investigate the CE for several distinct models in which the localised magnetic structure describe magnetic skyrmions and half-skyrmions. We first consider skyrmion-like configurations with integer $(1)$ topological charge, and then the case of half-skyrmions with semi-integer $(1/2)$ topological charge. 

\subsection{$Q=1$} 

Here we consider three different models, the two first ones studied before in \cite {BDR,JMM1} and in \cite{JMM2}, and the third one being a new model. In the two first cases, the new results concern the CE behavior, and in the third case the new results stand for the topological structure itself, and also for the CE behavior. 

\subsubsection{Skyrmion} 

The first model is defined by the potential 
\be\label{p1}
P(\phi)=\frac{1}{2(1-s)^2}(1-\phi^2)^2 ,
\ee
 where $s$ is a real parameter, $s\in[0,1)$. This model contains the minima $\bar{\phi}_{\pm}=\pm 1$ and the corresponding equation of motion admits the analytical solution 
\be
\label{ph1}
\phi_{s}(r)=\frac{1-r^{2/(1-s)}}{1+r^{2/(1-s)}}. 
\ee
We notice that $\phi_{s}(0)=1$ and $\phi_{s}(\infty)=-1$, irrespective of the value of $s$. This means that $s$ does not interfere in the topological charge, but it may modify the topological charge density, changing the distribution of the magnetization along the radial coordinate. To see how this works, we use the above solution \eqref{ph1} and take $\delta=0$ in the magnetization \eqref{M}: we depict the profile of the skyrmion in Fig. \ref{skyrmion}$(a)$, for  $s=0.3$ (left panel) and  $s=0.9$ (right panel). We notice that around the core $r\approx 0$, the magnetization is oriented in the positive $\hat{z}$ sense (blue arrows) and, as $r$ increases toward larger and larger values, the magnetization changes sense, pointing toward negative $\hat{z}$ (red arrows) as one reaches the frontier of the localized structure. In Fig. \ref{skyrmion}(d) we illustrate how the color varies as $r$ increases from zero to larger and larger values. We see from Fig. \ref{skyrmion}(a) that the $s$ parameter controls the internal distribution of magnetization, with the passage from blue $(+{\hat z})$ to red $(-\hat{z})$ being more or less smooth, depending of $s$ being closer to or farther from zero. 

\begin{figure}[t!]
\centering
\includegraphics[scale=0.4]{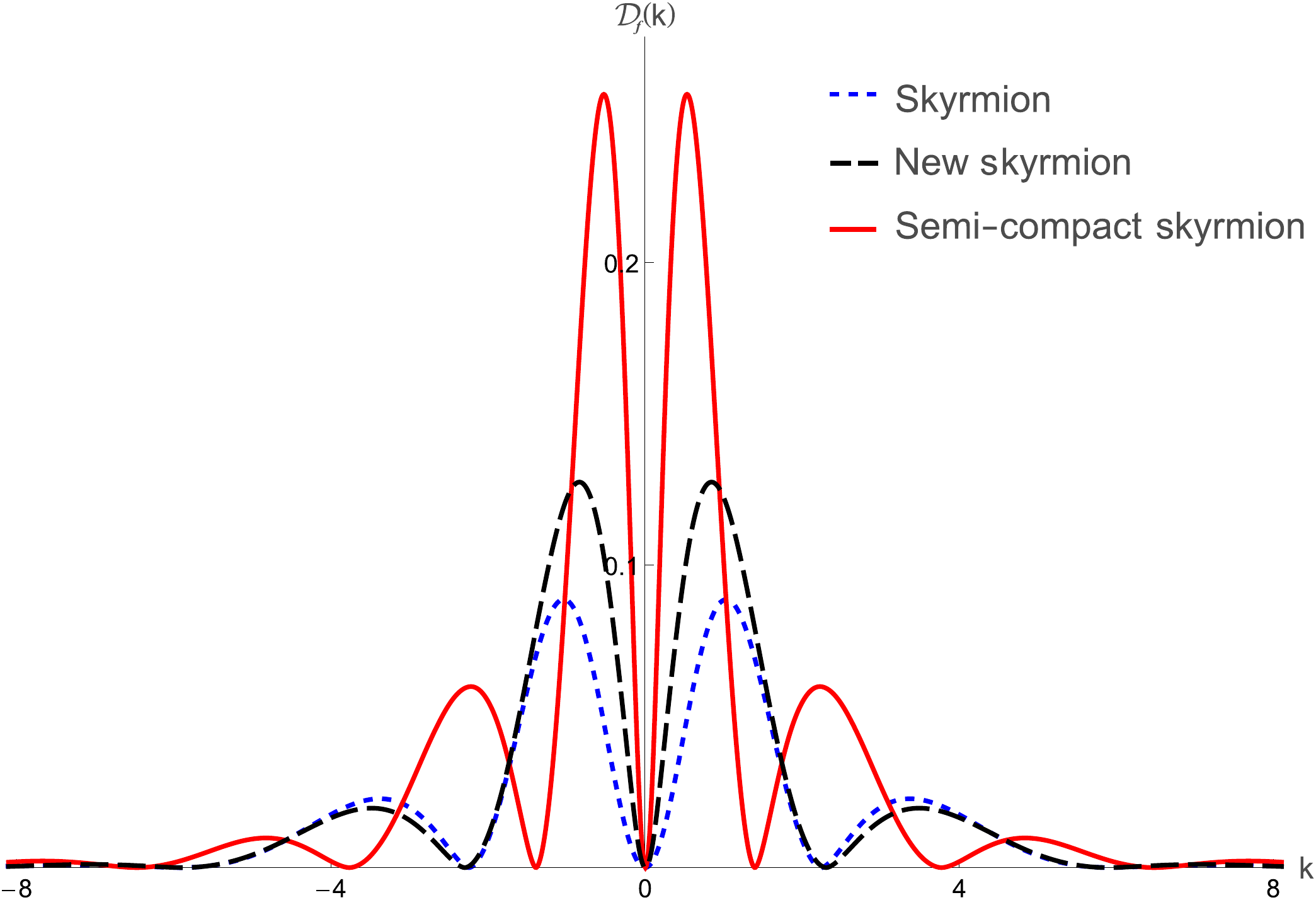}
\caption{The configurational entropy density ${\cal D}_f({\bf k}) $ is depicted with $s=0.3$, for the  models \eqref{p1} (dotted blue curve), \eqref{p2} (solid red curve) and \eqref{p5} (dashed black curve).}
\label{fig2}
\end{figure}

\begin{figure}[t!]
\centering
\includegraphics[scale=0.4]{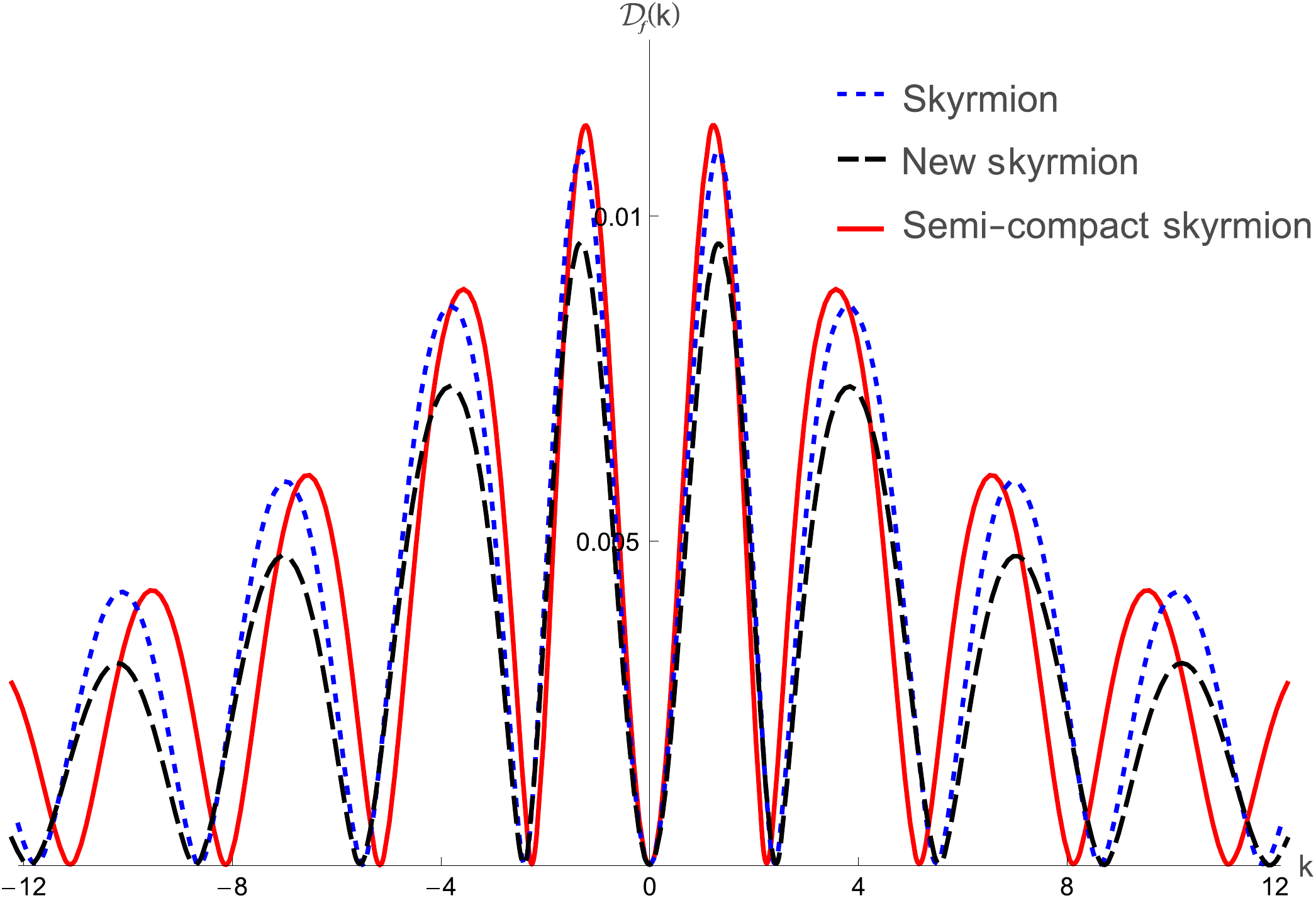}
\caption{The configurational entropy density ${\cal D}_f({\bf k}) $ is depicted with $s=0.9$, for the  models \eqref{p1} (dotted blue curve), \eqref{p2} (solid red curve) and \eqref{p5} (dashed black curve).}
\label{fig3}
\end{figure}

The fact that $s$ controls the magnetization inside the localised structure will the further used below to see how the distribution of magnetization inside the structure changes the CE of the magnetic skyrmion. In this model, the topological charge density \eqref{T} has the form

\be
\label{q1}
q_{s}(r)= \widetilde{q}(r) \cos\frac{\pi}{2} \left(\frac{1-r^{2/(1-s)}}{1+r^{2/(1-s)}}\right),  
\ee
where
\be
\widetilde{q}(r)=\frac{\pi r^{(1+s)/(1-s)}}{(1-s)(1+r^{2/(1-s)})}.
\ee
To investigate the CE, we first implement a numerical investigation of the modal fraction and the CE density. The results for the CE density are then shown in Figs. \ref{fig2} and \ref{fig3}, for $s=0.3$ and $s=0.9$, respectively. They are represented by the dotted blue curves that appear in these two figures. We have noticed an increase in the number of peaks as one increases $s$; also, the distribution of peaks is symmetric around $k=0$, and the two global maxima are ${\cal D}_f(k)_{max}=0.0886$ for $k= \pm1.0354$, in the case of $s=0.3$, and two global maxima ${\cal D}_f(k)_{max}= 0.0109$ for $k= \pm 1.2188$, in the case $s=0.9$. We also display in Fig.~\ref{fig4}, the behavior of $S_{C}$ for the present model \eqref{p1}; it is shown by the dotted blue curve. We notice that as $s$ increases, the entropic information decreases. This result seems to comply with the natural understanding, since for $s$ closer to unity, the magnetization conforms to a core region with positive magnetization surrounded by a ring with negative magnetization, thus leaving not too much room for other possibilities. For small values of $s$, however, both the core and ring regions diminish, leaving much more room for the magnetization to accommodate at values other than blue (positive) and red (negative). This behavior shows that for $s$ close to unity, the magnetic structures described above becomes a rigid bimagnetic core and ring structures. For readers interested in some numerical values for the CE, see Table~\ref{tab1}.

\subsubsection{Semi-compact skyrmion} 

Let us now consider another model, which was first studied in \cite{JMM2}. It is defined by 
\be\label{p2}
P(\phi)=\frac{1}{2(1-s)^2}(1+\phi)^2(1-\phi^n)^2 ,
\ee
where $s\in[0,1)$ and $n$ is a positive odd integer, $n=1,3,5,...$. The model support minima at $\bar{\phi}_{\pm}=\pm 1$ and engenders analytical solutions for $n=1$, which takes us back to the model \eqref{p1}. However, it is also exactly solved for $n\rightarrow \infty$:

\begin{eqnarray}
\label{phi2}
\phi_s(r)=\left\lbrace
\begin{array}{ll}
    \ 1, \quad 0 < r \leq 1; \\ 
    \ \\
    \ \displaystyle \left( \frac{2-r^{1/(1-s)}}{r^{1/(1-s)}}\right), \quad r > 1. \\ 
\end{array}
\right.
\end{eqnarray}
For other values of $n$, the solution can be obtained numerically. 

\begin{figure}[t!]
\centering
\includegraphics[scale=0.62]{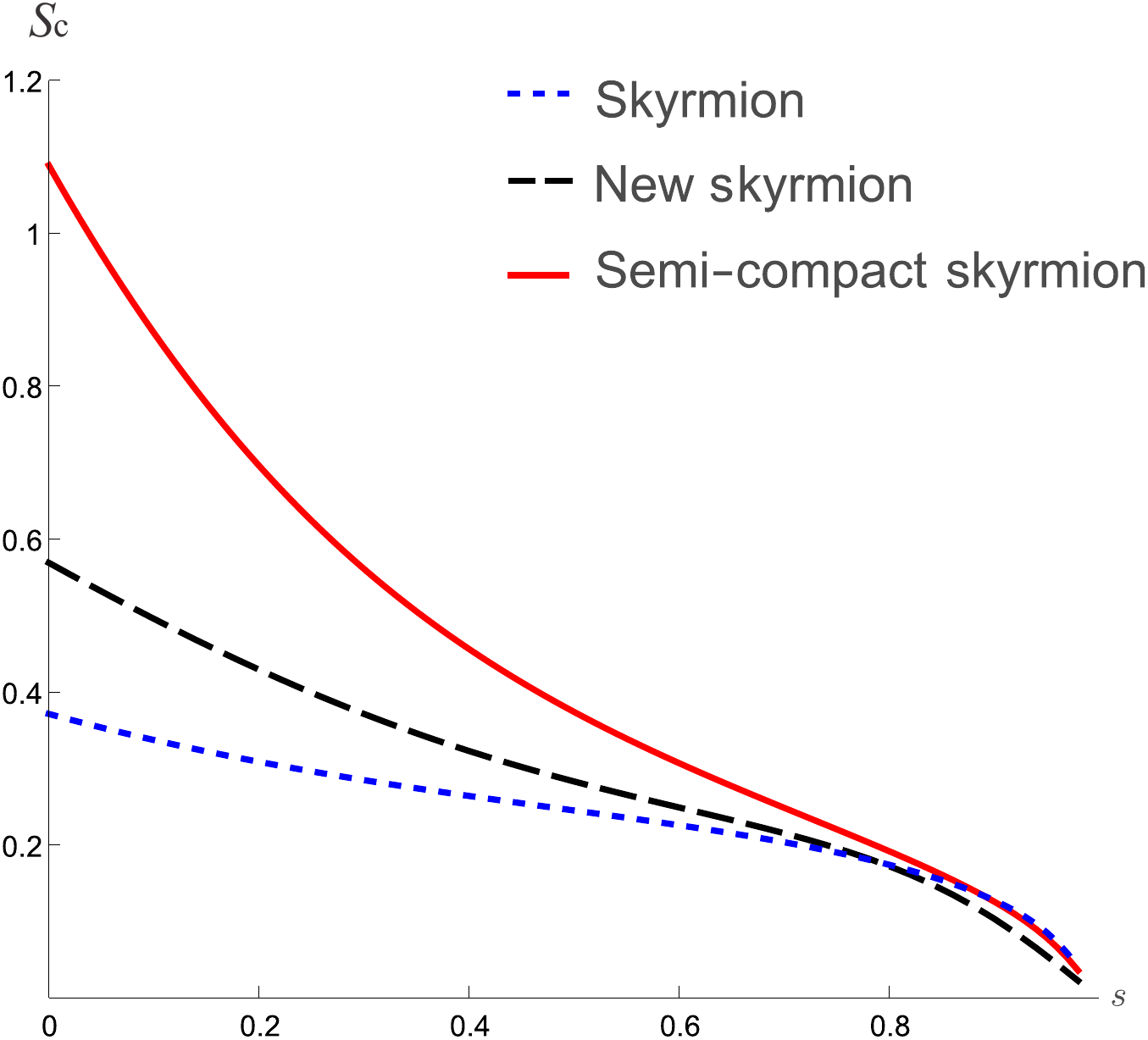}
\caption{The configurational entropy $S_c$ is depicted for the models \eqref{p1} (dotted blue curve), \eqref{p2} (solide red curve) and \eqref{p5} (dashed black curve).}
\label{fig4}
\end{figure}
The magnetization in Eq.~\eqref{M}, for $\delta=0$, combined with the above solution \eqref{phi2}, has a profile which is qualitatively similar to the case of the previous model; see Fig. \ref{skyrmion}(b), where the case with $n\to \infty$ is depicted for $s=0.3$ and $0.9$. We notice, however, that the abrupt change from $\hat{z}$ (blue) to $-\hat{z}$ (red) occurs for smaller values of $s$, when compared to the previous model. We also notice the same behavior identified before, that for $s$ close to unity, the magnetic skyrmion becomes a rigid bimagnetic core and ring structure.  

The topological charge density \eqref{T} for the present model has the form

\begin{eqnarray}
\label{q2}
q_s(r)=\left\lbrace
\begin{array}{ll}
    \ 0, \quad 0 < r \leq 1; \\ 
    \ \\
    \ \displaystyle \widetilde{q}(r) \cos\frac{\pi}{2} \left( \frac{2-r^{1/(1-s)}}{r^{1/(1-s)}}\right), \quad r > 1, \\ 
\end{array}
\right.
\end{eqnarray}
where
\be
\widetilde{q}(r)=\frac{\pi r^{-(2-s)/(1-s)}}{2(1-s)}.
\ee
We use this to numerically investigate CE density ${\cal D}_f(k)$ of the corresponding magnetic structure. The results are shown with solid red curves in Figs.~\ref{fig2} (for $s=0.3$) and \ref{fig3} (for $s=0.9$). They are qualitatively similar to the previous model: we also have an increase in the number of peaks as $s$ increases in the interval $[0,1)$, with symmetric distribution around $k=0$. The global maxima are ${\cal D}_f(k)_{max}=0.2556$ at $k= \pm 0.5295$, for $s=0.3$ and ${\cal D}_f(k)_{max}= 0.0012$ at $k= \pm 1.212$ for $s=0.9$. The corresponding CE is depicted in Fig.~\ref{fig4} with the solid red curve. It shows that the CE decreases more importantly, when compared to the previous model \eqref{p1}, depicted with the dotted blue curve. These two red and blue curves cross each other at $s=0.89$, where $S_C=0.1337$. See also Table 1 for some numerical values of the CE.


\subsubsection{New skyrmion} 

The third model is described by the potential

\be\label{p5}
P(\phi)=\frac{1}{2(1-s)^2}(1-2|\phi|+\phi^2).
\ee
This is a new model in the context of planar magnetic strutures. It was already investigated with other motivations in Refs.~\cite{ABL, BIL}. The potential has two minima at $\bar{\phi}_{\pm}=\pm 1$, and the solution 
\be
\label{phi5}
\phi_{s}(r)=\frac{\left(1-r^{1/(1-s)}\right)\left(e^{-\left|g(r)\right|}+1\right)}{1+r^{1/(1-s)}},  
\ee
where $g(r)=\log(r^{1/(1-s)})$.

The magnetization for $\delta=0$ is shown in Fig.\ref{skyrmion}$(c)$; For $s=0.3$ (left panel), we see that the magnetization changes sense slowly, from positive $\hat{z}$ (blue arrows) to negative $-\hat{z}$ (red arrows). For $s=0.9$ the change is more abrupt, but it is smoother when compared with the two previous models. We also notice the same behavior identified before, that for $s$ close to unity, the magnetic skyrmion becomes a rigid bimagnetic core and ring structure. 

In this model, the topological charge density \eqref{T} is combined with \eqref{phi5} to give
\be
\label{q5}
q_{s}(r)= \widetilde{q}(r) \cos\left[ \frac{\pi\left(1-r^{1/(1-s)}\right)\left(1+e^{-\left|g(r)\right|}\right)}{2(1+r^{1/(1-s)})}\right],  
\ee
where
\begin{eqnarray}
\widetilde{q}(r)&=& \frac{\pi r^{s/(1-s)}e^{-|g(r)|}}{2(1-s)(1+r^{1/(1-s)})|g(r)|} \left[ \left|g(r)\right|\left(1+e^{\left|g(r)\right|}\right) \right. \nonumber \\ 
&+& \left. \frac{ r^{1/(1-s)}}{2}g(r)\left(  r^{1/(1-s)}g(r) -1\right)\right].
\end{eqnarray}
The corresponding ${\cal D}_f(k)$, Eq.~\eqref{dsk}, is depicted with dashed black curves in Figs.~\ref{fig2} and \ref{fig3}, for $s=0.3$ and $s=0.9$, respectively. In a way similar to the two previous model, here we also see the increase in the number of peaks as $s$ increases in the interval $[0,1)$. We also have two global maxima ${\cal D}_f(k)_{max}=0.1274$ at $k= \pm 0.8414$, for $s=0.3$, and two others ${\cal D}_f(k)_{max}= 0.094$ at $k= \pm 1.3121$, for $s=0.9$. For the CE, we integrate numerically ${\cal D}_f(k)$ and depict the result in Fig.~\ref{fig4} with the dashed black curve. It is always smaller than the red curve of the model \eqref{p2}, and greater than the dotted blue curve of the model \eqref{p1} until $s=0.7885$, where the dashed black curve crosses the dotted blue curve. At this $s$ the CE is $S_{C}=0.1776$. See also Table 1, for other numerical values for the CE.

\begin{table*}[t!]
\centering
     \small
     \caption[]{The configurational entropy $S_c$ for some values of $s$ for the six models with $Q=1$ and $Q=1/2$.}
\begin{tabular}{lrrrrrrr}
\hline
\hline
\begin{minipage}[t]{.07\textwidth}\begin{flushleft} $ $\end{flushleft}\end{minipage}&
\begin{minipage}[t]{.13\textwidth}\begin{flushright}\ \ \ \ \ \ \ \ \ \ \ \   Skyrmion\end{flushright}\end{minipage}&
\begin{minipage}[t]{.13\textwidth}\begin{flushright} \ \ \ \ \ \ \ \ New skyrmion \end{flushright}\end{minipage}&
\begin{minipage}[t]{.13\textwidth}\begin{flushright} Semi-comp. skyrmion \end{flushright}\end{minipage}&
\begin{minipage}[t]{.13\textwidth}\begin{flushright}\ \ \ \ \ Half-skyrmion\end{flushright}\end{minipage}&
\begin{minipage}[t]{.13\textwidth}\begin{flushright} New half-skyrmion \end{flushright}\end{minipage}&
\begin{minipage}[t]{.13\textwidth}\begin{flushright}Semi-comp. half-skyrmion \end{flushright}\end{minipage}\\ 
\hline
$s=0.0$ &$0.3713$	&$0.5689$ &$1.0885$  &$0.7452$ &$0.8540$   &$0.9152$  \\
\hline
$s=0.3$ &$0.2847$	&$0.3713$ &$0.5610$  &$0.4131$ &$0.4687$  &$0.4995$ \\
\hline
$s=0.6$ &$0.2257$	&$0.2490$ &$0.3070$  &$0.2644$ &$0.2599$  &$0.2927$ \\
\hline
$s=0.9$ &$0.1276$	&$0.1037$ &$0.1260$  &$0.1567$ &$0.1781$  &$0.1125$ \\
\hline
\hline
\end{tabular}
\label{tab1}
\end{table*}

\begin{figure}[t!]
\centering
\includegraphics[scale=0.57]{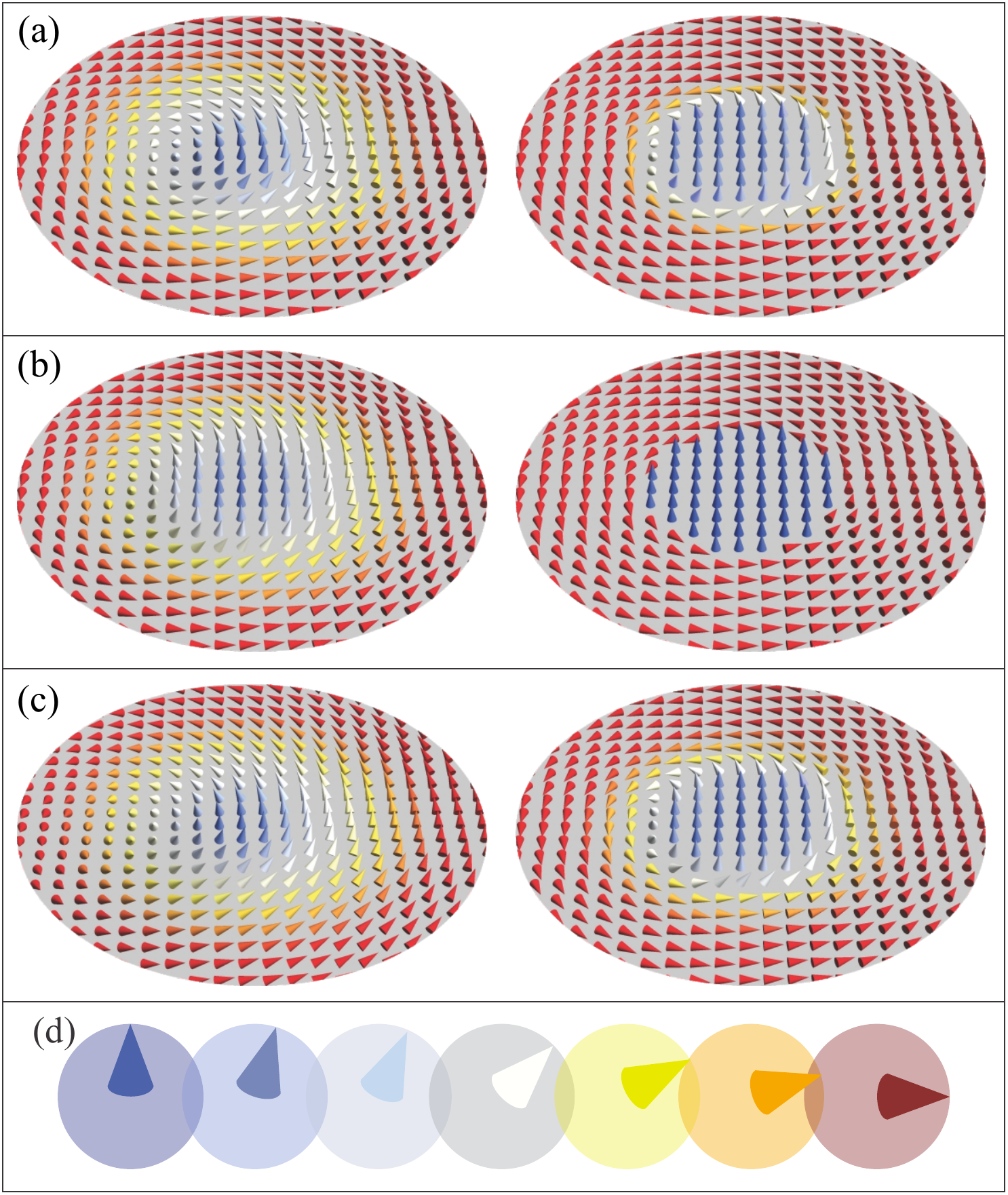}
\caption{The magnetic structures with $Q=1/2$ for (a) model \eqref{p3}, (b) model \eqref{p4} and (c) model \eqref{p6}, depicted for $s=0.3$ (left) and $s=0.9$ (right). The bottom panel in (d) illustrates how the magnetization ${\bf m}(0)=\hat{z}$ (blue) changes to $\hat{\theta}$ (red) as $r$ increases to larger and larger values.}
\label{half_skyrmion}
\end{figure}

\subsection{$Q=1/2$} 

Let us now investigate the CE associated to three distinct models that attain topological charge $Q=1/2$. We first deal with two models studied before in Refs. \cite{BDR,JMM1} and \cite{JMM2}, with distinct motivations, and then move on and consider a new model. In the three cases, we present new results for the CE, which are controlled by the magnetization of the corresponding localized structures.

\subsubsection{Half-skyrmion} 

We start with the model defined by the potential

\be\label{p3}
P(\phi)=\frac{1}{2(1-s)^2} \phi^2(1-\phi^{2})^2. 
\ee
It has minima at $\bar{\phi}_{\pm}=\pm 1$ and $\bar{\phi}_{0}=0$, and maxima at $\phi_{max}^{\pm}=\pm 1/ \sqrt{3}$, with $P(\phi_{max}^{\pm})=2/(27(1-s)^{2})$. The solution that connects the minima $0$ and $1$ is given by

\be
\phi_{s}(r)=\frac{r^{1/(1-s)}}{(r^{2/(1-s)} + 1)^{1/2}}.  
\ee
It also obeys the first order equation \eqref{ordeq}. We notice that $\phi_{s}(0)=0$ and $\phi_{s}(\infty)=1$, and the magnetization \eqref{M} for $\delta = \pi/2$ gives ${\bf M}(0)=\hat{z}$ and ${\bf M}(\infty)=\hat{\theta}$. This is further shown in Fig. \ref{half_skyrmion}$(a)$ for $s=0.3$ (left panel) and $s=0.9$ (right panel). We also notice the same behavior identified before, that for $s$ close to unity, the magnetic configuration becomes a rigid bimagnetic core and ring structure. 

\begin{figure}[h!]
\centering
\includegraphics[scale=0.4]{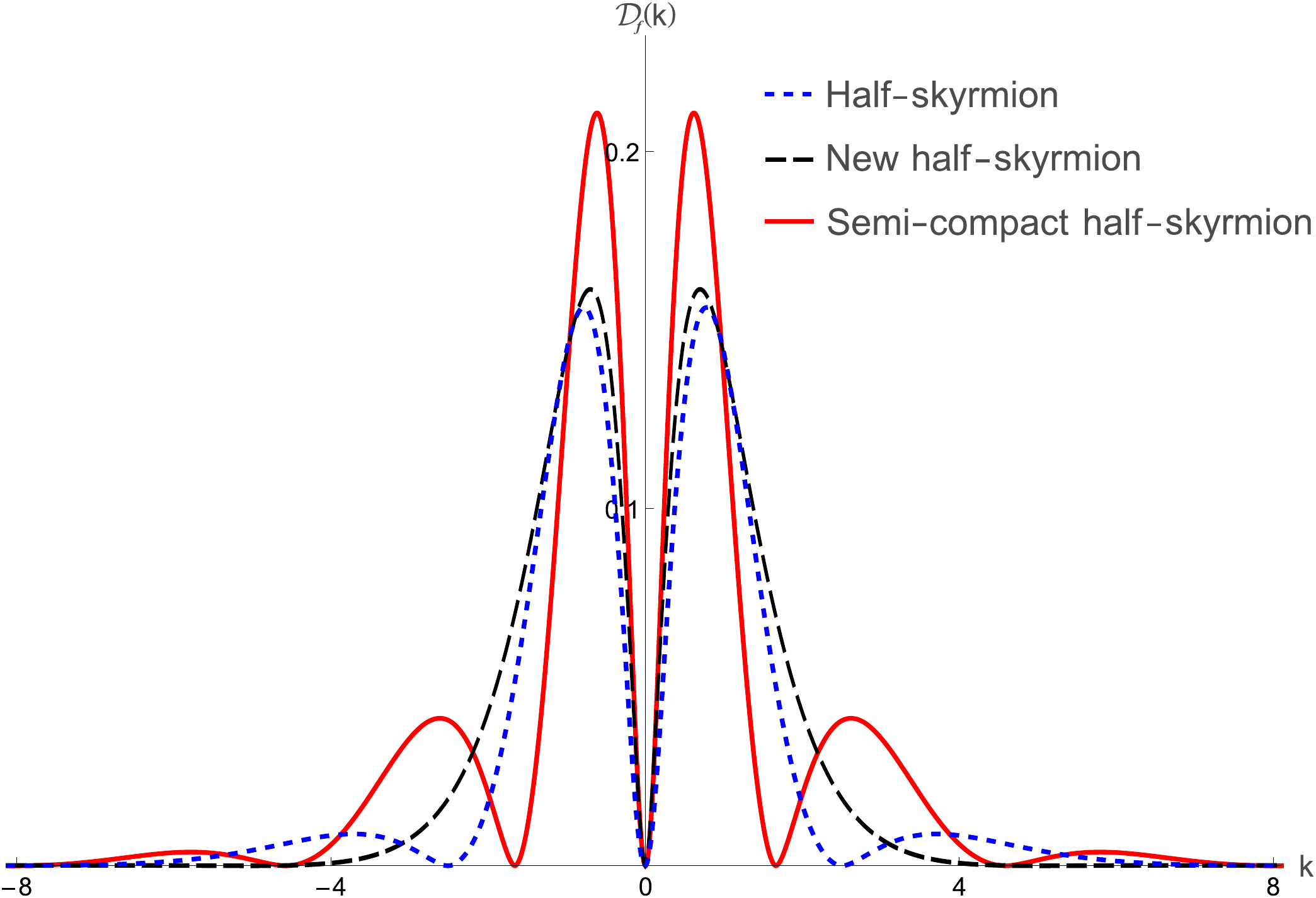}
\caption{The configurational entropy density ${\cal D}_f({\bf k}) $ is depicted with $s=0.3$, for the models \eqref{p3} (dotted blue curve), \eqref{p4} (solid red curve) and \eqref{p6} (dashed black curve).}
\label{fig6}
\end{figure}

\begin{figure}[h!]
\centering
\includegraphics[scale=0.4]{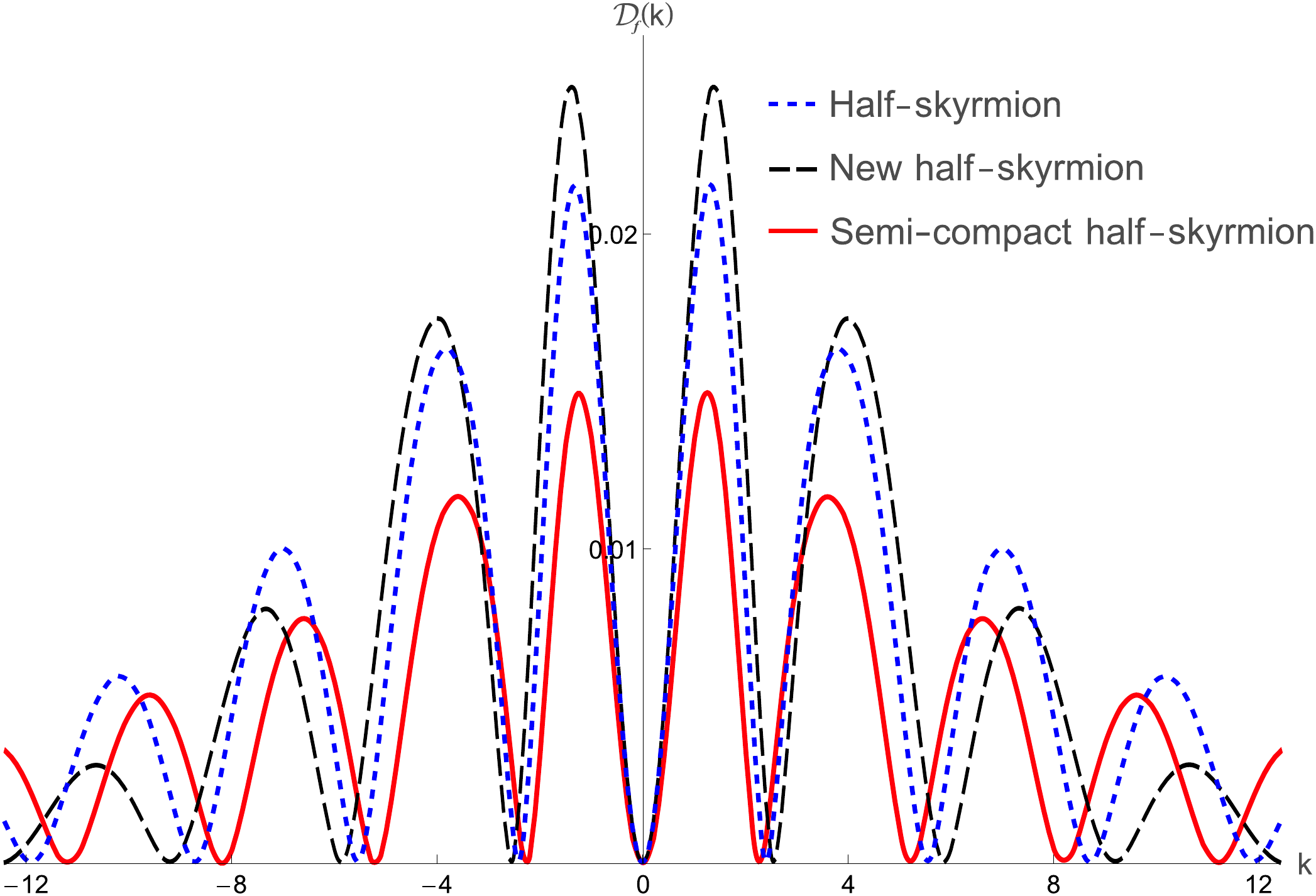}
\caption{The configurational entropy density ${\cal D}_f({\bf k}) $ is depicted with $s=0.9$, for the models \eqref{p3} (dotted blue curve), \eqref{p4} (solid red curve) and \eqref{p6} (dashed black curve).}
\label{fig7}
\end{figure}

In this model, the topological charge density has the form

\be
\label{q3}
q_{s}(r)= \widetilde{q}(r) \sin\frac{\pi}{2} \left(\frac{r^{1/(1-s)}}{(r^{2/(1-s)} + 1)^{1/2}}\right),  
\ee
where
\be
\widetilde{q}(r)=\frac{\pi r^{s/(1-s)}}{4(1-s)(1+r^{2/(1-s)})^{3/2}}.
\ee
We use the above result to numerically describe the density ${\cal D}_f(k)$ in Figs.~\ref{fig6} and \ref{fig7}, for $s=0.3$ and $s=0.9$, respectively. The results are depicted with the dotted blue curves, and they are qualitatively similar to the cases of skyrmions. Here we also identified two global maxima ${\cal D}(k)_{max}=0.1562$ at $k= \pm 0.7802$ for $s=0.3$, and two global maxima ${\cal D}(k)_{max}= 0.0172$ at $k= \pm 1.3216$ for $s=0.9$. The CE is displayed in Fig.~\ref{fig8} with the dotted blue curve, and one notices that it diminishes as we increase $s$ in the interval $[0,1)$. The behavior here is also qualitatively similar to the cases of skyrmions. We also present some numerical value of the corresponding CE in Table 1.

\subsubsection{Semi-compact half-skyrmion}

The second model is described by the potential 
\be\label{p4}
P(\phi)=\frac{1}{2(1-s)^2}\phi^2 \text{ln}^2(\phi^2).
\ee
It also has three minima, at $\bar{\phi}_{0}=0$ and $\bar{\phi}_{\pm}=\pm 1$, and two maxima at $\phi_{max}^{\pm}=\pm 1/e$ such that $P(\phi_{max}^{\pm})=(2e^{-2})/(1-s)^{2}$. The corresponding first order equation supports the solution 

\be\label{phi4}
\phi_{s}(r)= e^{-r^{-2/(1-s)}},  
\ee
which connects the minima $\bar{\phi}_{0}=0$ and $\bar{\phi}_{+}=1$. We use \eqref{phi4} and \eqref{M} with $\delta=\pi/2$ to see the magnetic structure in Fig. \ref{half_skyrmion}$(b)$ for $s=0.3$ (left panel) and $s=0.9$ (right panel). The behavior is similar to the case of the semi-compact skyrmion already investigated. In fact, the change in the magnetization from $\hat{z}$ (blue arrows) to $\theta$ (red arrows) is more abrupt for larger values of $s$ in the interval $[0,1)$. We also notice the same behavior identified before, that for $s$ close to unity, the magnetic half-skyrmion becomes a rigid bimagnetic core and ring structure.

In this model, the topological charge density has the form

\be
\label{q4}
q_{s}(r)=\widetilde{q}(r) \sin\frac{\pi}{2} \left(e^{-r^{-2/(1-s)}}\right),  
\ee
where
\be
\widetilde{q}(r)=\frac{\pi r^{-(3-s)/(1-s)}e^{-r^{-2/(1-s)}}}{2(1-s)}.
\ee
We use this to numerically investigate the configurational entropy density ${\cal D}_f(k)$. The results are displayed in Figs.~\ref{fig6} and \ref{fig7}, for $s=0.3$ and $s=0.9$, respectively. We notice that ${\cal D}_f(k)$ engenders two global maxima given by ${\cal D}_f(k)_{max}=0.2107$ at $k= \pm 0.6168$, for $s=0.3$ and two others ${\cal D}_f(k)_{max}= 0.0124$ at $k= \pm 1.2449$, for $s=0.9$. We also depict the CE $S_{C}$ in Fig.~\ref{fig8} with the solid red curve. The red solid curve cross the blue dotted curve at $s=0.7568$, where $S_{C}=0.2170$. For some numerical values of the CE, see Table 1. \\


\subsubsection{New half-skyrmion} 

The third model was first studied in Ref.~\cite{BIL}, where the authors investigated stability of the solution and the thermal effects due to the potential. Here we study the CE behavior associated to this model, which is defined by

\be\label{p6}
P(\phi)=\frac{1}{2(1-s)^2}\phi^2(1-2|\phi|+\phi^2).
\ee
There are minima at $\bar{\phi}_{0}=0$ and $\bar{\phi}_{-}=\pm1$, and maxima at $\phi_{max}^{\pm}=\pm 1/2$, where $P(\phi_{max}^{\pm})=3/(32(1-s)^{2})$. In the sector connecting $0$ and $1$, the solution is 

\be
\label{phi6}
\phi_{s}(r)=\frac{r^{1/(1-s)}}{1+r^{1/(1-s)}}.
\ee

\begin{figure}[h!]
\centering
\includegraphics[scale=0.62]{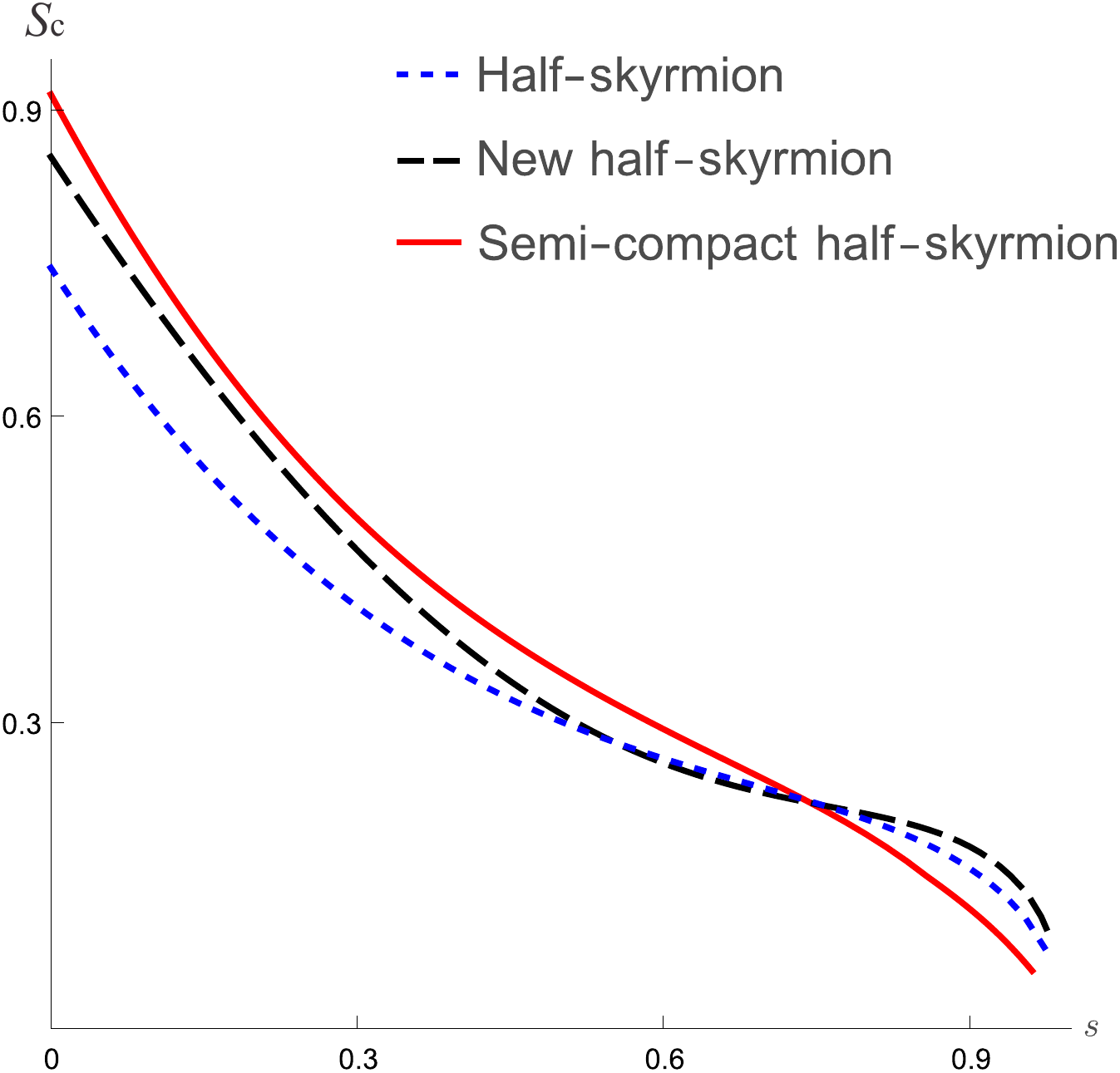}
\caption{The configurational entropy $S_c$ is depicted for the models \eqref{p3} (dotted blue curve), \eqref{p4} (solide red curve) and \eqref{p6} (dashed black curve).}
\label{fig8}
\end{figure}

We use the magnetization \eqref{M} with $\delta = \pi/2$ and the above solution to see how it varies as one changes $s$. The results are displayed in Fig. \ref{half_skyrmion}$(c)$, for $s=0.3$ (left panel) and for $s=0.9$ (right panel). Here we also notice the same behavior identified before, that for $s$ close to unity, the magnetic skyrmion tends to become a rigid bimagnetic core and ring structure. We then use the Eq.~\eqref{T} and the solution \eqref{phi6} to get the topological charge density

\be
\label{q6}
q_{s}(r)= \widetilde{q}(r)\sin\left[ \frac{\pi}{2}\left(\frac{r^{1/(1-s)}}{1+r^{1/(1-s)}}\right)\right],  
\ee
where
\begin{eqnarray}
\widetilde{q}(r)=\frac{\pi  r^{s/(1-s)}}{4(1-s)(1+r^{1/(1-s)})^2}.
\end{eqnarray}
The associated CE density ${\cal D}_f(k)$ is depicted in Figs.~\ref{fig6} and \ref{fig7} with dashed black curves for $s=0.3$ and $s=0.9$, respectively. We see that ${\cal D}_f(k)$ has two global maxima ${\cal D}_f(k)_{max}=0.1620$ at $k= \pm 0.7017$ for $s=0.3$, and two others ${\cal D}_f(k)_{max}= 0.0247$ at $k= \pm 1.3874$ for $s=0.9$. The CE $S_C$ in this case is depicted in Fig.~\ref{fig8} with the dashed black curve. It reaches the dotted blue curve at $s=0.5519$, where $S_{C}=0.2814$. For some numerical values of the CE, see Table 1.

\section{Comments and conclusions}

In this work we studied the configurational entropy (CE) associated to several localized magnetic structures that attain topological behavior with integer $Q=1$ and half-integer $Q=1/2$ charge, which are usually called skyrmions and half-skyrmions, respectively. We investigated three distinct systems that support skyrmion-like structures, and three other systems that support half-skyrmion-like structures. 

The behavior of the magnetization inside the structures are qualitatively similar but quantitatively different, and the difference appears in each one of the several CE profile. The magnetization depends on the parameter $s\in[0,1)$ that controls each one of the six models, and its behavior in between the core and the ring regions is smoother or steeper, depending on $s$ being closer to zero or unity, respectively. For $s$ increasing toward unity, the core and ring regions increase, leaving almost no room for other arrangements; the magnetic arrangements tend to become a rigid bimagnetic core and ring structure, with the CE decreasing as $s$ increases. On the reverse, for $s$ decreasing toward zero, the core and shell regions decrease, leaving more and more room for other magnetic arrangements, increasing the CE toward higher values. This is the general behavior, and it appears for both skyrmions and half-skyrmions, although they give quantitatively different results for each specific case.

The topological charge of skyrmions and half-skyrmions are fixed at integers and half-integers, but the results found in this work show that the internal disposition of magnetization is also important. It may lead to topological structures capable of engendering much more room for the disposition of information, and this is an issue of current interest. In the models studied in this work, we have added a single parameter capable of controlling the internal disposition of magnetization. In other scenarios, in the models investigated in \cite{S1,S2,S3,S4,S4a,S5,BS}, for instance, the authors consider other possibilities, and we hope the present work may foster new investigations on this and other related issue. 

The fact that the smoother distribution of magnetization inside the magnetic structure leads to richer entropic content does not imply that rigid distributions of magnetization are not of current interest. For instance, bimagnetic elements with rigid core and shell structures are currently being studied with a diversity of applications. An interesting motivation, in particular, is the possibility to provide much more precise measurements when subjected to appropriate magnetic probes. In this sense, bimagnetic nanometric core and shell elements may be useful to enhance diagnostic imaging outputs as they appear in magnetic resonance imaging; see, e.g., \cite{review,book} and references therein for further information on this and for other applications of current interest.

\acknowledgments{This research is supported in part by Conselho Nacional de Desenvolvimento Cient\'\i fico
e Tecnol\'ogico (CNPq, Grants Nos. 404913/2018-0 and 303469/2019-6) and by Para\'\i ba State Research Foundation (FAPESQ-PB, Grant No. 0015/2019).}


\end{document}